\documentclass{article}

\usepackage{arxiv}

\usepackage[utf8]{inputenc} 
\usepackage[T1]{fontenc}    
\usepackage{hyperref}       
\usepackage{url}            
\usepackage{booktabs}       
\usepackage{amsfonts}       
\usepackage{nicefrac}       
\usepackage{microtype}      
\usepackage{lipsum}
\usepackage{graphicx}
\usepackage{setspace}
\usepackage{bm}
\usepackage{caption}
\usepackage{amsmath}
\usepackage{tabu}
\usepackage{amssymb}
\usepackage[authoryear]{natbib}
\renewcommand{\l}{\left}
\renewcommand{\r}{\right}
\newcommand{\R}{\mathbb{R}}

\graphicspath{ {./images/} }

\title{Come Together: Analyzing Popular Songs Through Statistical Embeddings}

\author{
 Matthew Esmaili Mallory \\
  Harvard University\\
  Department of Statistics\\
  Cambridge, MA, USA \\
  \texttt{matthewmallory@fas.harvard.edu} \\
   \And
 Mark Glickman \\
  Harvard University\\
  Department of Statistics\\
  Cambridge, MA, USA \\
  \texttt{glickman@g.harvard.edu} \\
  \And
 Jason Brown \\
  Department of Mathematics and Statistics\\
  Dalhousie University\\
  Halifax, NS, Canada \\
  \texttt{Jason.Brown@Dal.Ca} \\
}

\begin{document}
\setstretch{1.1}
\maketitle
\begin{abstract}
Statistical modeling of popular music presents a unique challenge due to the complexity of song structures, which cannot be easily analyzed using conventional statistical tools. However, recent advances in data science have shown that converting non-standard data objects into real vector-valued embeddings enables meaningful statistical analysis. In this work, we demonstrate an approach based on logistic principal component analysis to construct embeddings from global song features, allowing for standard multivariate analysis.  We apply this method to a corpus of Lennon and McCartney songs from 1962-1966, using embeddings derived from chords, melodic notes, chord and pitch transitions, and melodic contours. Our analysis explores how these song embeddings cluster by Beatles album, how songwriting styles evolved over time, and whether Lennon and McCartney's compositions exhibited convergence or divergence. This embedding-based approach offers a powerful framework for statistically examining musical structure and stylistic development in popular music.
\end{abstract}


\section{Introduction}\label{intro}

In the past few decades, there has been a growing interest in the study of music and songwriting through an analytical framework. Such work includes learning important features of various musical genres \citep{sham2017scratch}, applying tools from computational topology to understand voice leading \citep{bergomi2015phd}, and analyzing the harmonic structures of pop music \citep{burgoyne2013comp}. Generally, the statistical modeling of music can prove to be uniquely difficult, as the structure of a song is not immediately appropriate for common models in statistics or machine learning. There are also many characteristics of an individual song that can be understood in differing contexts, such as the sentiment of the lyrics, complexity of the instrumentation, or recording quality. As a result, no universally accepted method exists with which to convert and subsequently analyze musical data.

However, such a problem is not unique to music, as similar problems occur in the processing of photographs and video, text, graph networks, and other non-standard data. A popular approach to addressing this issue is to convert the data into real-valued \textit{embeddings} through methods such as dimension reduction or neural networks. The main idea behind embeddings is that they should capture the semantic connections between groups of points through their pairwise distances in Euclidean space. A common approach for constructing such embeddings is principal component analysis (PCA), in which one aims to project multivariate data into a lower-dimensional subspace while preserving as much variation as possible. For text data, popular machine learning algorithms like word2vec and BERT convert textual data into embeddings for two main purposes: next-word prediction and contextual understanding \citep{word2vec, BERT}. The ability to accurately model text data in this way has led to an explosion in the field of natural language processing (NLP), resulting in large language models such as ChatGPT and Gemini \citep{chatgpt, gemini}.

Our goal is to illustrate the relevance of embeddings for the purpose of converting and analyzing musical data. In many applications, the characteristics of a piece of music can be represented as binary variables indicating the presence or absence of certain features \citep{binaryfeatures2002, glickman2019data}. However, attempting to work directly with these binary features can prove to be difficult, especially when the data is high-dimensional and/or correlated. Converting the features into lower-dimensional real embeddings offers many advantages, such as reducing multicollinearity and allowing the use of popular statistical models that often require real-valued, continuous data. This enables performing meaningful analyses on the progression of musical style over time, comparisons between songwriters, authorship attribution, and much more. In this paper, we develop a method for creating embeddings of songs based on vectors of binary features that can be used for subsequent analysis. 

The outline for the rest of the paper is as follows. In Section \ref{methods}, we introduce Logistic Principal Component Analysis (Logistic PCA) as the main tool for creating embeddings \citep{landgraf2020pca, lee2010sparse, schein2003glm}. Section \ref{data} then describes how to apply these methods to a specific dataset consisting of a corpus of Beatles songs, along with performing several different analyses on these embeddings. We then present a possible musical explanation of our findings in Section \ref{disc} and discuss limitations and potential extensions.

\section{PCA for Binary Data}\label{methods}

Classical PCA is a widely used tool for dimension
reduction in multivariate data that are assumed to be approximately (multivariately) normally distributed, and can be used as the first step in constructing embeddings for observations.
The fundamental idea is to project data into a lower-dimensional space which preserves as much variation as possible. 
However, with multivariate binary data, classical PCA is not directly applicable. 
Several recent works
\citep{landgraf2020pca, collins2001efpca, deleeuw2006pca, schein2003glm} 
have developed procedures to apply PCA principles to non-Normal and/or binary data.

Suppose that we observe a data matrix $\bm X \in \{0, 1\}^{n\times d}$ consisting of $n$ observations of $d$ binary features. In the context of musical analysis, we might have
\begin{align}
    x_{ij} = \begin{cases}
        1, & \text{ the } j^{\text{th}} \text{ musical feature is present in song } i ,\\
        0, & \text{otherwise.}
    \end{cases}
\end{align}
If we model each entry as $x_{ij} \sim \text{Bern}(p_{ij})$, then the corresponding 
real-valued natural parameters in an exponential family framework
are $\theta_{ij} := \text{logit}(p_{ij})$. 
Prior work \citep{collins2001efpca, schein2003glm, deleeuw2006pca} 
generalizes the methods of PCA to exponential families by considering the parameter matrix $\bm\Theta := (\theta_{ij}) \in \R^{n\times d}$, 
and positing a low-rank decomposition of the form 
\begin{align}\label{eq:expfampca}
    \bm\Theta = \bm 1_n\bm{\mu}^\intercal + \bm A\bm B^\intercal,
\end{align}
where $\bm 1_n = (1, \hdots, 1)^\intercal$ is the vector of all ones, and $\bm{\mu}$ is a vector of feature-specific intercepts (or main effects). Here, $\bm A$ and $\bm B$ are expected to be of a much lower rank $k \ll \min(n, d)$, which can represent $k$ latent factors. This is very similar to a classical factor analysis: in our case, the rows of $\bm A \in \R^{n\times k}$ correspond to length-$k$ vectors describing the latent profile of each song, and the columns of $\bm B^\intercal \in \R^{k \times d}$ summarize how strongly each feature loads onto each of these latent factors. Such a low-rank structure means that we now only estimate $(n+d)k + d$ parameters instead of $nd$, which can be much less for small values of $k$. 

Our primary focus, however, is on logistic PCA, as presented in \cite{landgraf2020pca}. We briefly outline their main procedure below.
Rather than explicitly factorizing the natural parameter matrix, logistic PCA proceeds by projecting the parameters from the Bernoulli saturated model to a lower-dimensional space, in a way that minimizes the Bernoulli deviance
\begin{align}
    \mathcal D(\bm\Theta \mid \bm X) &=-2\sum_{i, j}
    \left(\vphantom{\sum_{i=1}^n}
    x_{ij}\log(p_{ij}) + (1 - x_{ij})\log(1-p_{ij})\right)\nonumber
    \\ &= -2\sum_{i, j}
    \left(\vphantom{\sum_{i=1}^n}
    x_{ij}\log\sigma(\theta_{ij}) + (1 - x_{ij})\log\sigma(-\theta_{ij})
    \right),
\end{align}
where $\sigma(t) := (1+e^{-t})^{-1}$ is the sigmoid function. 

An important consideration is that, for the Bernoulli saturated model, the Bernoulli deviance is minimized by setting $p_{ij} = x_{ij}$. This results in the saturated natural parameters $\tilde\theta_{ij}$ being infinite:
\begin{align}
    \tilde\theta_{ij} = \begin{cases}
        -\infty, & x_{ij} = 0\\
        +\infty, & x_{ij} = 1.
    \end{cases}
\end{align}
To be computationally feasible, $\tilde\theta_{ij}$ is instead restricted to $\pm m$ for a fixed $m > 0$ chosen later. Denoting the matrix of these bounded natural parameters by $\tilde{\bm\Theta}$, the optimization problem associated with logistic PCA can be written as 
\begin{align}
    \min_{\substack{\bm U^\intercal \bm U = I_k \\ \bm{\mu} \in \R^d}}\mathcal D\left(\mathbf{1}_n\bm{\mu}^\intercal + (\tilde{\bm\Theta} - \mathbf{1}_n\bm{\mu}^\intercal)\bm U\bm U^\intercal \mid \bm X\right).
\end{align}
Notably, fixing $\bm{\mu}=\bm0$ as proposed by the authors, $\bm\Theta$ factors as $\tilde{\bm\Theta}\bm U\bm U^\intercal$ instead of the $\bm A \bm B^\intercal$ decomposition as in Equation \eqref{eq:expfampca}. This has the advantage of only requiring estimation of a single matrix, which is more tractable for high-dimensional data.

Based on the formulation described above, two main parameters must be determined as inputs to perform logistic PCA: the number of principal components, $k$, and the truncation parameter, $m$.
To choose $m$, one may perform row-wise cross-validation to determine an optimal value. 
Choosing a value of $k$ depends on the practitioner's use for their data. For visualization purposes, it is natural to choose $k \leq 3$. One may also compute the minimal $k$ for which 80\% or 90\% of the cumulative variance in the data matrix is explained, and use this value explicitly.

Logistic PCA has been implemented in the 
\texttt{logisticPCA} package in R \citep{landgraf2020pca}.
The package allows performing the cross-validation mentioned above to find $m$, and returns the principal component scores (of the form $\tilde{\bm\Theta}\bm U$) to use for analysis.

\section{Application to Musical Data}\label{data}

It is difficult to overestimate the impact that The Beatles had on the popular music industry and music as a whole. As a testament to their enduring legacy, they won the 2024 Grammy award for best rock performance over half a century after breaking up. The Beatles discography presents a wealth of opportunities for statistical analysis, whether that be authorship prediction, evolution of style over time, or understanding the complex dynamics that influence writers. 

The dataset we analyzed was first constructed in \cite{glickman2019data}, and is based on songs released by The Beatles between 1962 and 1966. This period consists of their first seven albums \textit{Please Please Me}, \textit{With the Beatles}, \textit{A Hard Day's Night}, \textit{Beatles For Sale}, \textit{Help!}, \textit{Rubber Soul}, and \textit{Revolver}, along with all the singles over the same time period. We included only the albums produced before The Beatles quit touring in late 1966, as subsequently they began to dive into more complex, experimental studio production and considerably altered their sound. Each observation corresponds to a single song, and consists of the album, authorship, and 137 binary features indicating the presence or absence of various harmonic and melodic features.

The rest of this section proceeds as follows. In Section \ref{datadescription}, we describe the dataset, including an explanation of the various musical features. In Section \ref{logpcaimp} we detail how logistic PCA is implemented in R. In Section \ref{convandvar} we utilize this model to explore how the musicality of Lennon and McCartney evolved, both numerically and through visualizations. We then compare the songwriting characteristics of George Harrison to Lennon and McCartney in Section \ref{other_artists}. Finally, in Section \ref{clusters} we compare the predicted authorship for a set of disputed Beatles songs across different models, including logistic regression on the logistic PCA components, $k$-nearest neighbors (KNN) classification, random forest, and the results of \cite{glickman2019data}. 

\subsection{Dataset Description}\label{datadescription}

\begin{figure}[!t]
    \centering
    \includegraphics[width=1\linewidth]{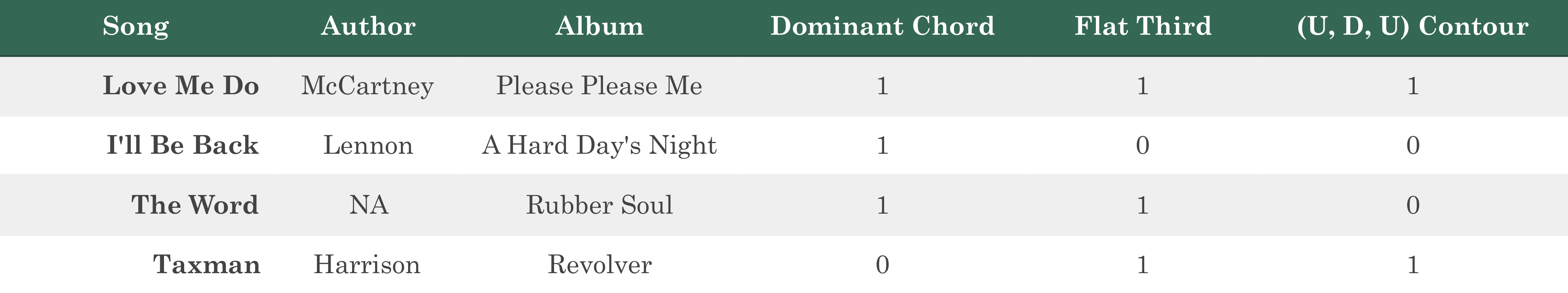}
    \caption{A small sample of the Beatles dataset, showing each song's authorship, album, and a few features. ``Dominant Chord,'' ``Flat Third,'' and ``(U, D, U) Contour'' represent the presence of the dominant chord, the flat third, and an \textit{up, down, up} contour, respectively.}
    \label{fig:data-table}
\end{figure}

We describe below a brief overview of the dataset. A complete description of the dataset along with a detailed explanation of feature construction can be found in \cite{glickman2019data}. 

Each song consists of an associated binary vector $\bm{X}_i \in \{0, 1\}^{137}$ indicating the presence of various musical features. When constructing these features, five main categories were considered: pitches, chords, pitch transitions, harmonic transitions, and contours. 
As in \cite{glickman2019data}, the melodic and harmonic features of songs were 
standardized in relation to the tonic/key of the song, so that in effect every song was treated as being in the same key (or relative minor key).

The first category, pitches, refers to the occurrence of different pitch classes in the chromatic scale at some point in a song. As we do not consider the octave of the note, there are exactly twelve pitches that could occur within the melody of a song.

The second category, chords, refers to which chord types are present in the song. Since they are so prevalent in western popular music, the occurrence of seven diatonic chords (I, ii, iii, IV, V, vi, vii${}^{\text{o}}$) are assigned their own features. However, since the diminished chord vii${}^{\text{o}}$ is relatively uncommon, it is replaced with the more standard minor vii chord. Then all non-diatonic major chords and all augmented chords were grouped into one feature, and all non-diatonic minor chords and diminished chords were also grouped into one feature. Finally, all extended chords (seventh, ninth, eleventh) were grouped with their respective triads. \citet{glickman2019data} considered other categorizations, but ultimately settled on this approach, as leaving less popular chord types on their own can lead to overly sparse feature vectors and thus unreliability in the final model. 

The third category, pitch transitions, indicates the presence of sequential pairs of notes in the melody of the song. In contrast to the single pitch occurrences, \citet{glickman2019data} did consider the octave of just the second note, which results in a three-octave range. Since this leads to $12 \cdot 36 = 432$ possible sequential pairs, various transitions are aggregated into a smaller number of groupings. The exact pairings are described in detail in \cite{glickman2019data}. Examples of these groupings include a repetition of any non-diatonic note, a phrase ending on the tonic, and moving down a semi-tone from diatonic to non-diatonic.

The fourth category, harmonic transitions, indicates the presence of consecutive chord pairs within a song. Since the tonic, subdominant, and dominant chords are so prevalent, any movement between them was its own feature, such as $\text{I} \to \text{V}$ or $\text{IV}\to\text{I}$. Similar to the pitch transitions, however, most of the category must otherwise be aggregated. This includes features such as moving between two non-diatonic chords or moving between a non-diatonic chord and the dominant chord.

The fifth and final category, contours, is substantially different from the other four. For each subset of four sequential notes in a melodic line within a song, the directionality of all three pairs of notes (moves up, down, or stays the same) is recorded, yielding $3^3 = 27$ total features. The use of longer local contours was also considered, but ultimately decided against as it would again lead to large dimensionality.

After including all of these features, our dataset $\bm X$ is of size $90 \times 137$, consisting mostly of Lennon and McCartney songs, with a few songs written by Harrison and a few that are either jointly written, unknown, or disputed.

In Figure \ref{fig:data-table}, we present a few observations and features. An important characteristic of our dataset is that it does not encode any information about the lyrical content, instrumentation, or audio itself. This is intentional, as we wished to focus only on the musical content as an approach to understanding songwriting without the external influence of the lyrical sentiment, recording process, and more. 

        

\subsection{Logistic PCA Implementation}\label{logpcaimp}

To choose the truncation parameter $m$, we performed cross-validation with the \texttt{cv.lpca()} function in R. Since $\sigma(10) \approx 1 - 10^{-5}$, with high accuracy it suffices to only consider $m \leq 10$. Larger values of $m$ correspond to fitted probabilities closer to the endpoints (0 and 1), and smaller $m$ corresponds to probabilities closer to 0.5. Ranging over the values $m \in \{1, 2, \hdots, 10\}$, we performed the cross-validation and achieved the highest likelihood for $m = 3$. 
This value of $m$ was used in all subsequent analyses. 

Next, to choose $k$, we determined the cumulative proportion of variation explained as a function of the number of principal components. In our case, it turned out that 35 principal components could recover 80\% of the variation in the data. Based on this analysis, we used 35 principal components (PCs) as sufficient representation of the songs. 

\subsection{Convergence and Variation in Songwriting Over Time}\label{convandvar}

\begin{figure}[!t]
    \centering    \includegraphics[width=0.85\linewidth]{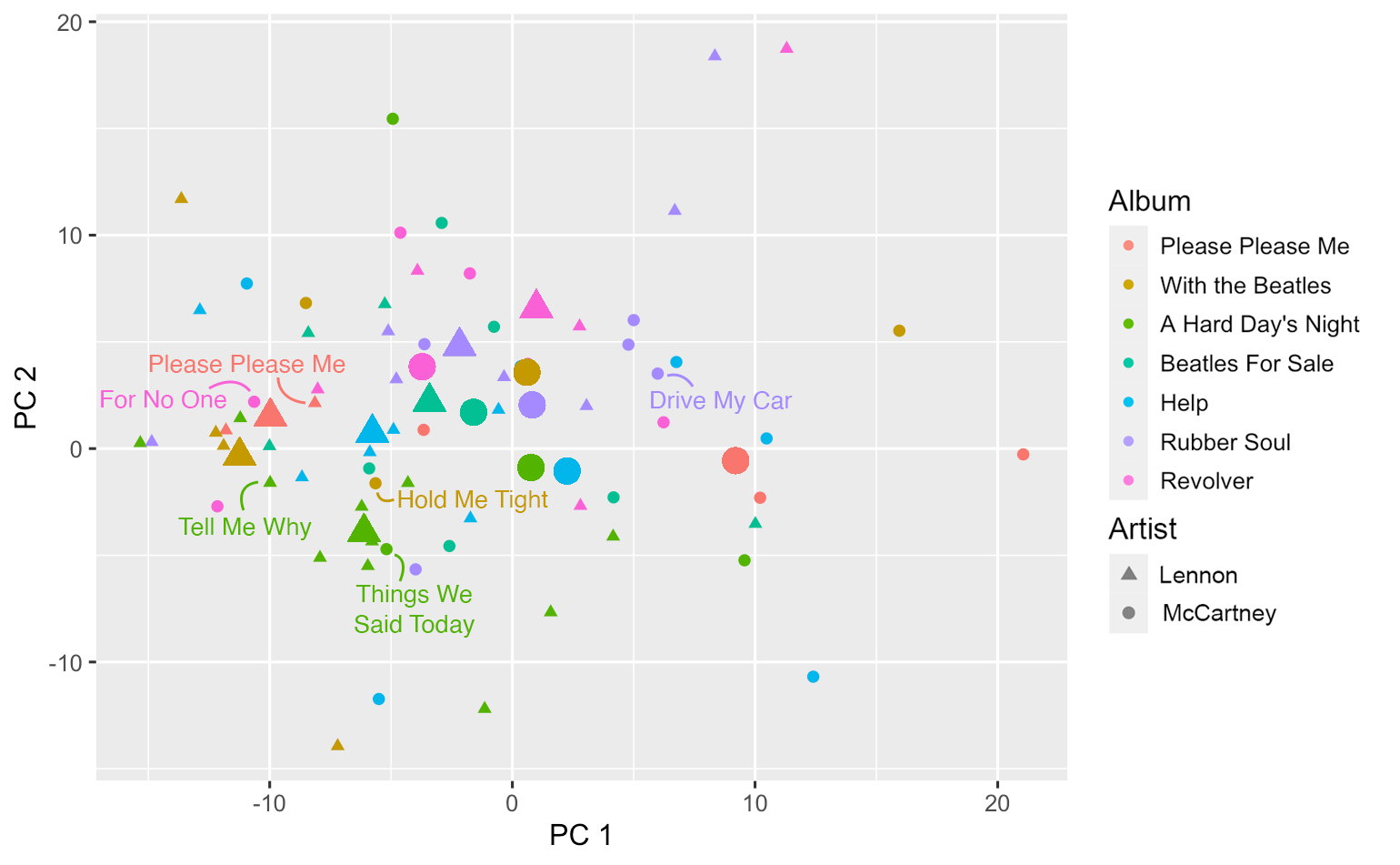}
    \caption{Plot of the first two principal components, colored by album. Circles represent McCartney’s songs, triangles represent Lennon’s songs, and enlarged points are album centroids. The six labeled songs are those identified as outliers by the OGK algorithm.}
    \label{fig:pcs-with-interaction}
\end{figure}

\begin{figure}[!t]
    \centering
        
    \includegraphics[width=0.8\linewidth]{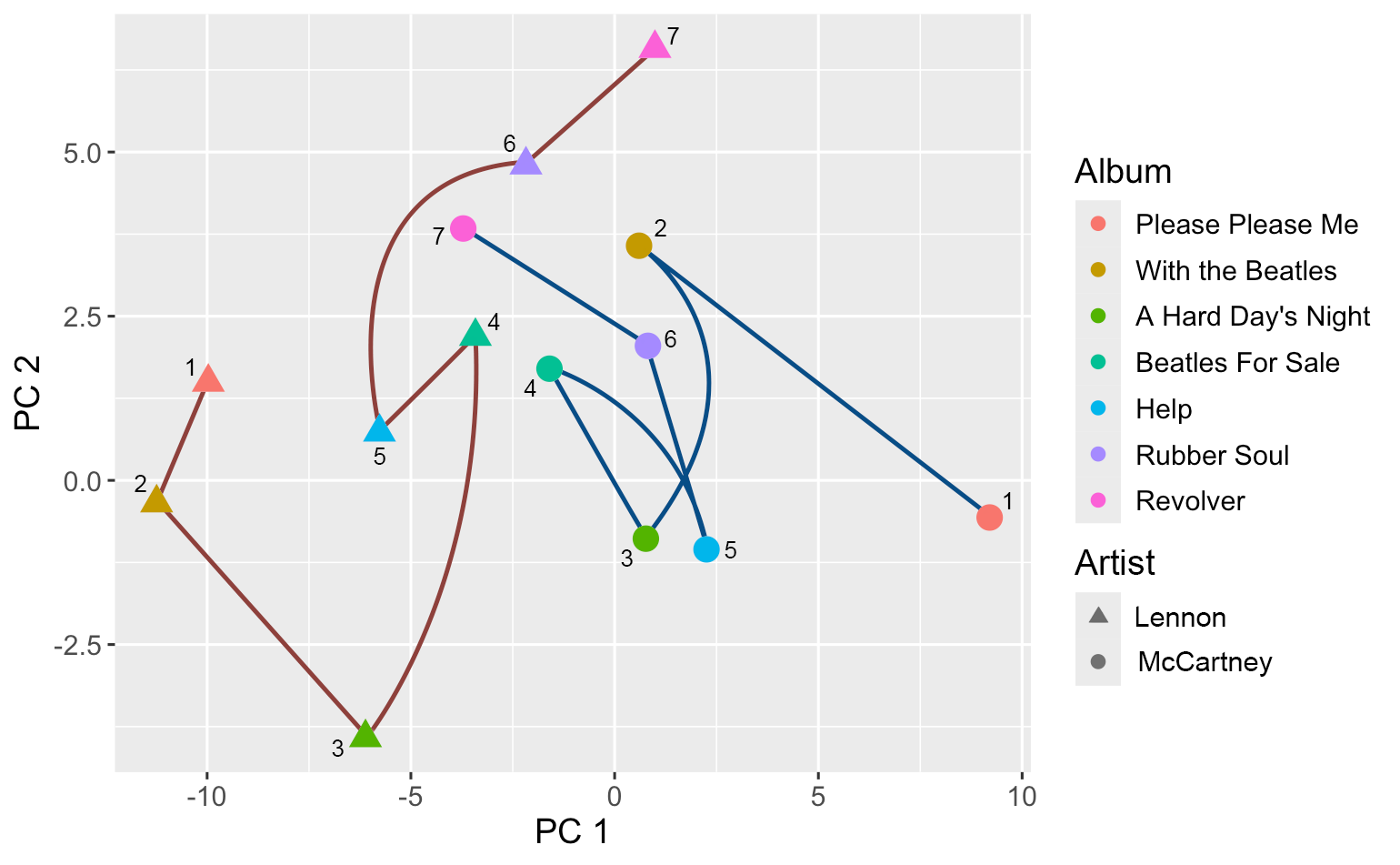}
    \caption{Plot of the album centroids for the first two principal components, again colored by album. Circles represent McCartney’s songs, triangles represent Lennon’s songs, and lines track the movement of each songwriter’s album-wise centroids.}
    \label{fig:centroids-only}
\end{figure}

We now explore the ways in which the styles of Lennon and McCartney
converged, diverged, and varied throughout their creative collaboration, offering a unique perspective on musical authorship and stylistic development.

First, we visualize the evolution of Lennon and McCartney's songwriting by projecting song embeddings onto their first two principal components, grouped by the album and/or artist. These figures highlight both the convergence of their styles over time and the trajectory of their within-album variability. It is worth noting that the first two principal components account for only about 12\% of total deviance, so the plots of the first two PCs are useful for visualization purposes only. Our subsequent numerical analyses were conducted based on the full 35 PCs.

In Figure \ref{fig:pcs-with-interaction}, we graph the first two PCs colored by album and author, along with their respective album centroids. There is quite a substantial scatter across the plot, and along the first principal component, Lennon and McCartney album centroids appear to be nearly split across $\text{PC}_1 = 0$. Their centroids also seem to converge toward one another across time, moving from the extremes of the graph toward the middle. We have also labeled the six songs identified as outliers by the Orthogonalized Gnanadesikan-Kettenring (OGK) algorithm, which we explain more in detail in Section \ref{disc}.

In Figure \ref{fig:centroids-only}, we simplify this graph by only examining the album centroids for Lennon and McCartney and sequentially tracking their movement with segments between albums. Similar to the previous graph, we observe that both songwriters' centroids begin on opposite endpoints of the graph and gradually converge toward each other in the center. Furthermore, McCartney's centroids seem to be more tightly clustered than those of Lennon, whose centroids seem to shift more erratically, especially with regard to $\text{PC}_2$. 

\begin{figure}[!t]
    \centering
        
    \includegraphics[width=0.8\linewidth]{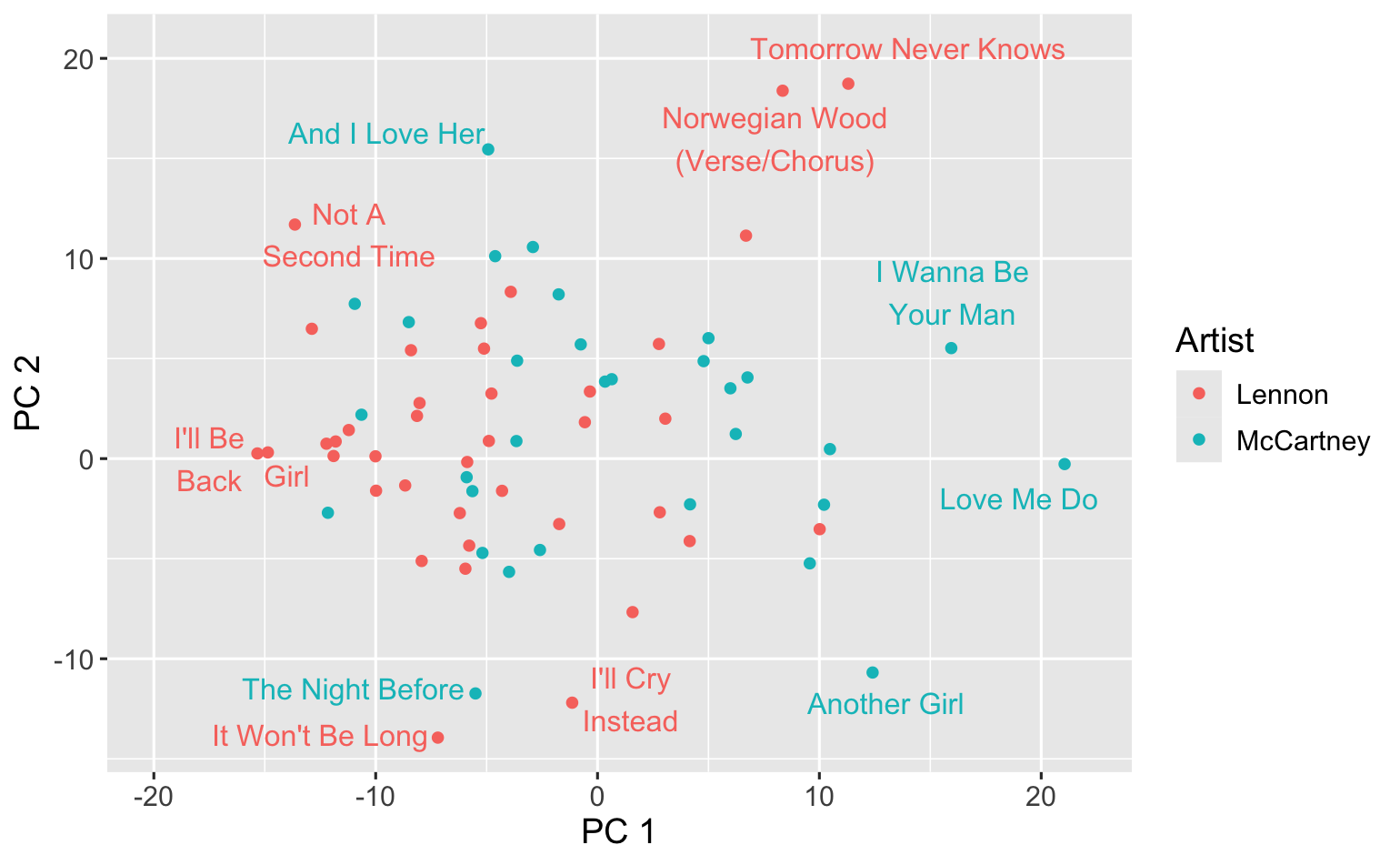}
    \caption{Plot of the first two principal components for songs by Lennon and McCartney.}
    \label{fig:new-interaction}
\end{figure}

In Figure \ref{fig:new-interaction}, we again plot the first two principal components, but this time only colored by artist (Lennon and McCartney). The purpose of this visualization is to understand which songs by each author could be considered outliers based only on these first two principal components. There are quite a few notable outliers by both songwriters, such as \textit{Tomorrow Never Knows} and \textit{Norwegian Wood} for Lennon, and \textit{Love Me Do} and \textit{Another Girl} for McCartney. However, due to the first two principal components explaining less than 20\% of the deviance in authorship, we exclude discussing any plausible explanations for these outliers, and exclude that musical discussion to only the outliers found by the OGK algorithm in Section \ref{disc}. 

\begin{figure}[!t]
    \centering
        
    \includegraphics[width=0.8\linewidth]{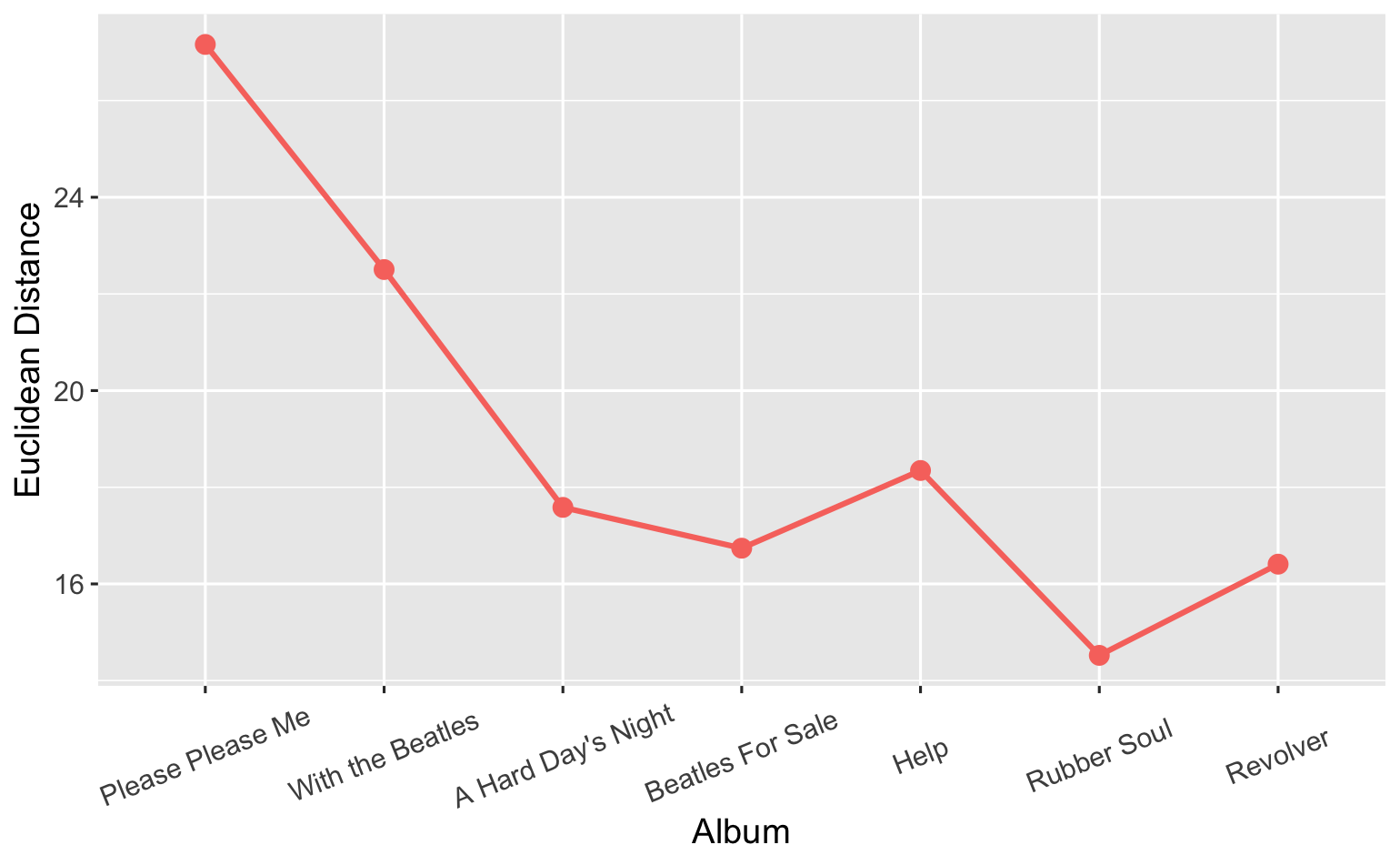}
    \caption{Average Euclidean distance between Lennon’s and McCartney’s embeddings across all albums from 1962 to 1966. The decreasing trend indicates that their musical styles became more similar over time.}
    \label{fig:album-diffs}
\end{figure}

We also examined these musical embeddings numerically, allowing us to understand whether Lennon and McCartney's songwriting characteristics became more or less similar over time. 
The following analyses are based on 35 PCs.

In Figure \ref{fig:album-diffs}, we plot the Euclidean distance between the centroids of Lennon- and McCartney-authored songs for each of the albums from \textit{Please Please Me} to \textit{Revolver}. Similar to the previous graphs, we observe a general trend of decreasing distance over time. The drop is quite sharp for the first few albums, and then slightly flattens toward the end. 

\begin{figure}[!t]
    \centering
        
    \includegraphics[width=0.8\linewidth]{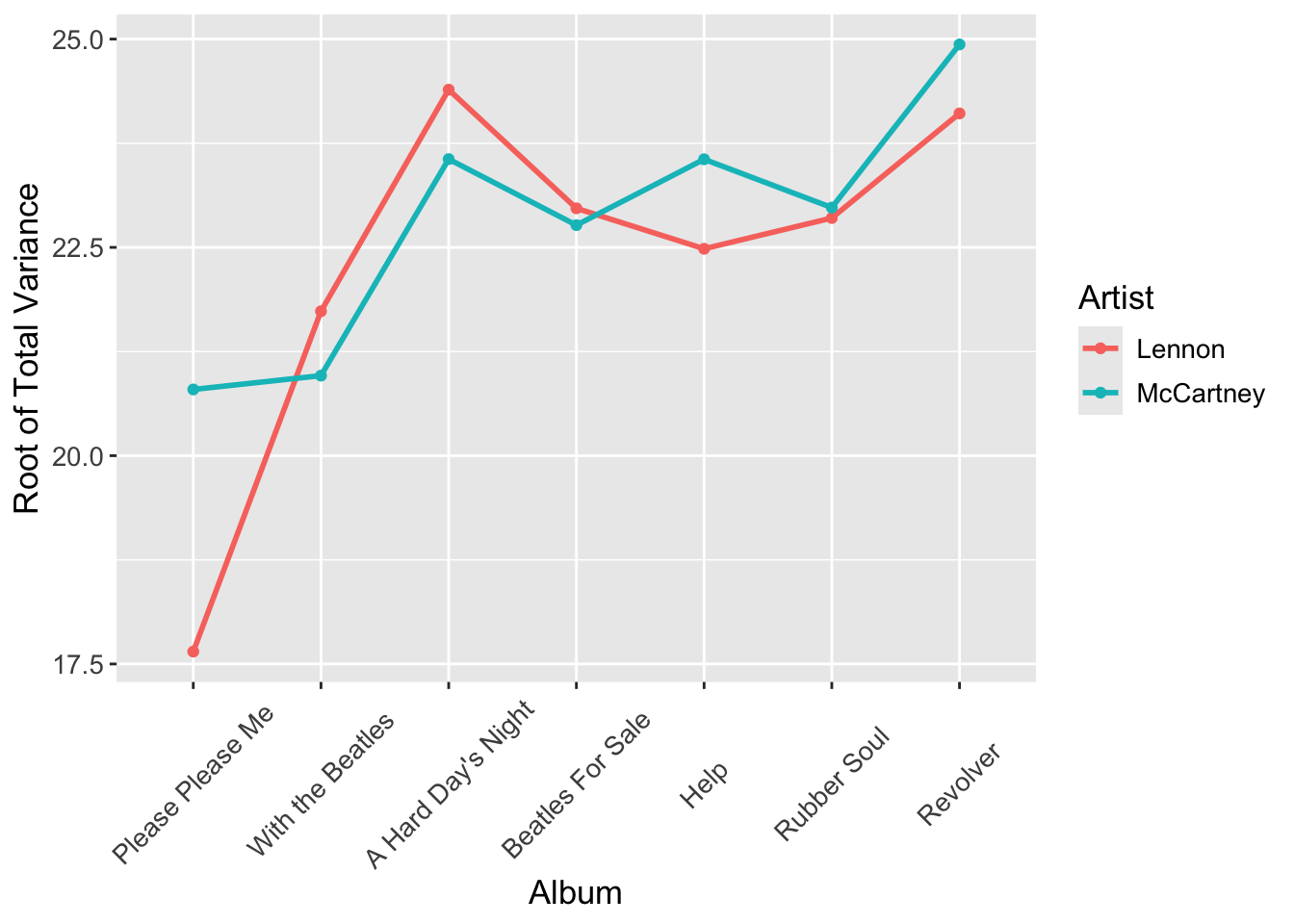}
    \caption{Plot of the square root of total variance, which is computed as the average squared Euclidean distances from Lennon’s and McCartney’s embeddings to their respective album centroids. Both artists show increasing stylistic variance over time, following a similar trajectory.}
    \label{fig:dist-to-cent}
\end{figure}

A similar visualization can be seen in Figure \ref{fig:dist-to-cent}, in which we plot the
square root of the average squared Euclidean distances from the centroids within each album for both artists, that is, the square root of the total variance. Once more, the two songwriters have a very similar trend of increasing variance over time. This trend is quite dramatic during the first few albums, and begins to flatten out but stay relatively high from \textit{A Hard Day's Night} and on.

\subsection{Stylistic Comparison to George Harrison}\label{other_artists}

\begin{figure}
    \centering
        
    \includegraphics[width=0.8\linewidth]{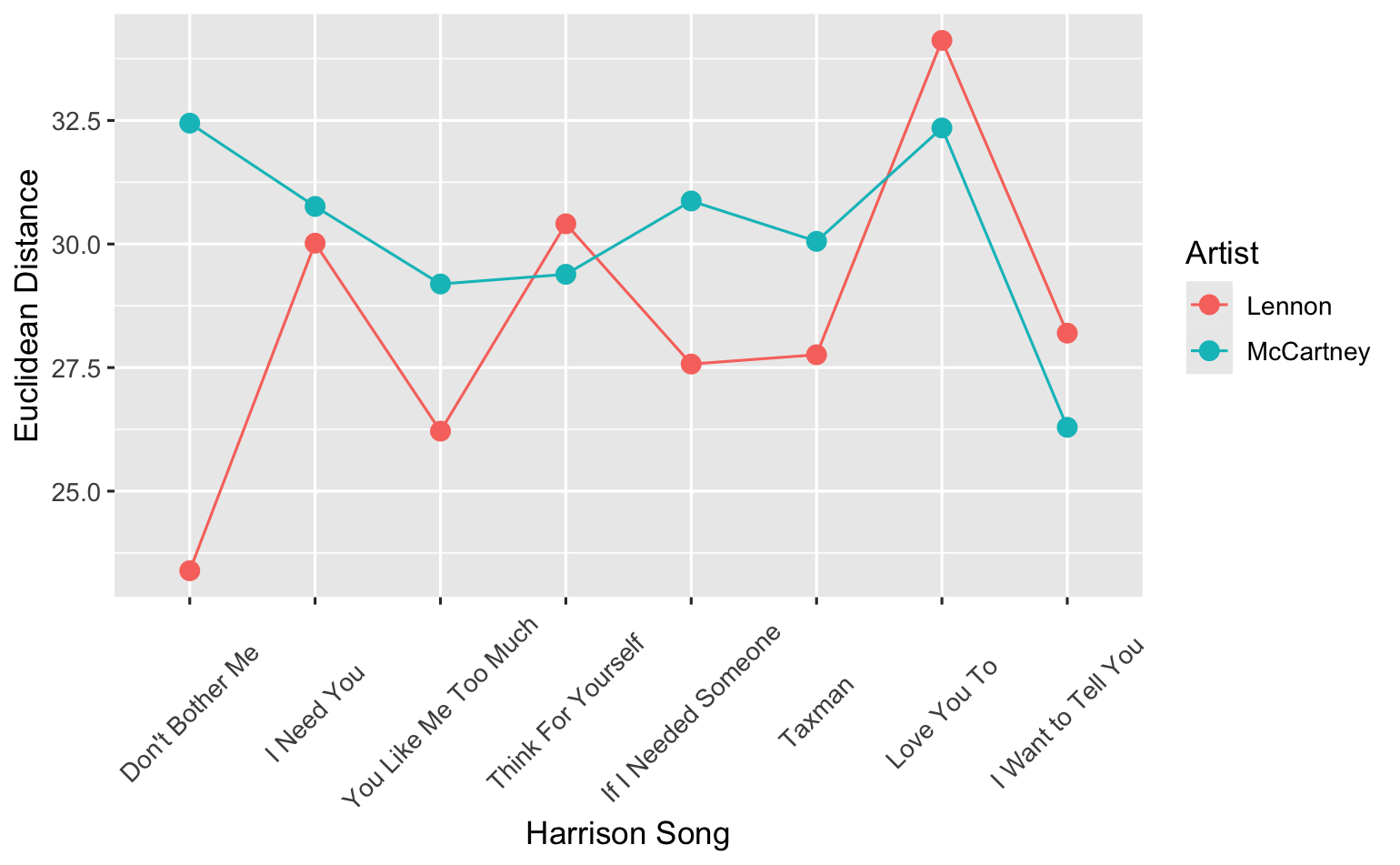}
    \caption{Plot showing the Euclidean distance between each of George’s songs and the corresponding album-specific centroids for Lennon and McCartney.}
    \label{fig:quietbeatle}
\end{figure}

One way we can also utilize these musical embeddings is to consider the features of a song or artist not explicitly used in constructing the principal components, and analyze how their song characteristics compare to that of our original list of songwriters. For our Beatles data, this easily lends itself to analyzing whether George Harrison, who also contributed several original songs during the 1962-66 period,
was more similar to Lennon or McCartney, and how this might have evolved over time.

Figure \ref{fig:quietbeatle} visualizes the distance from the embeddings of Harrison's eight songs to the album centroids of Lennon and McCartney from their respective albums. Notably, Harrison's distance to the McCartney centroids stays relatively constant over time, whereas the distance to the Lennon centroids fluctuates more substantially. Additionally, neither comparison shows a clear tendency for this similarity to increase or decrease over time. We discuss potential musical explanations for these phenomena in Section \ref{disc}. 

\subsection{Authorship Attribution}\label{clusters}

We consider here the problem of authorship prediction based on our song embeddings. As a first approach, we applied $k$-means clustering with $k=2$ on our principal components, aiming to distinguish between Lennon and McCartney songs.

\begin{figure}
    \centering    
    \includegraphics[width=0.80\linewidth]{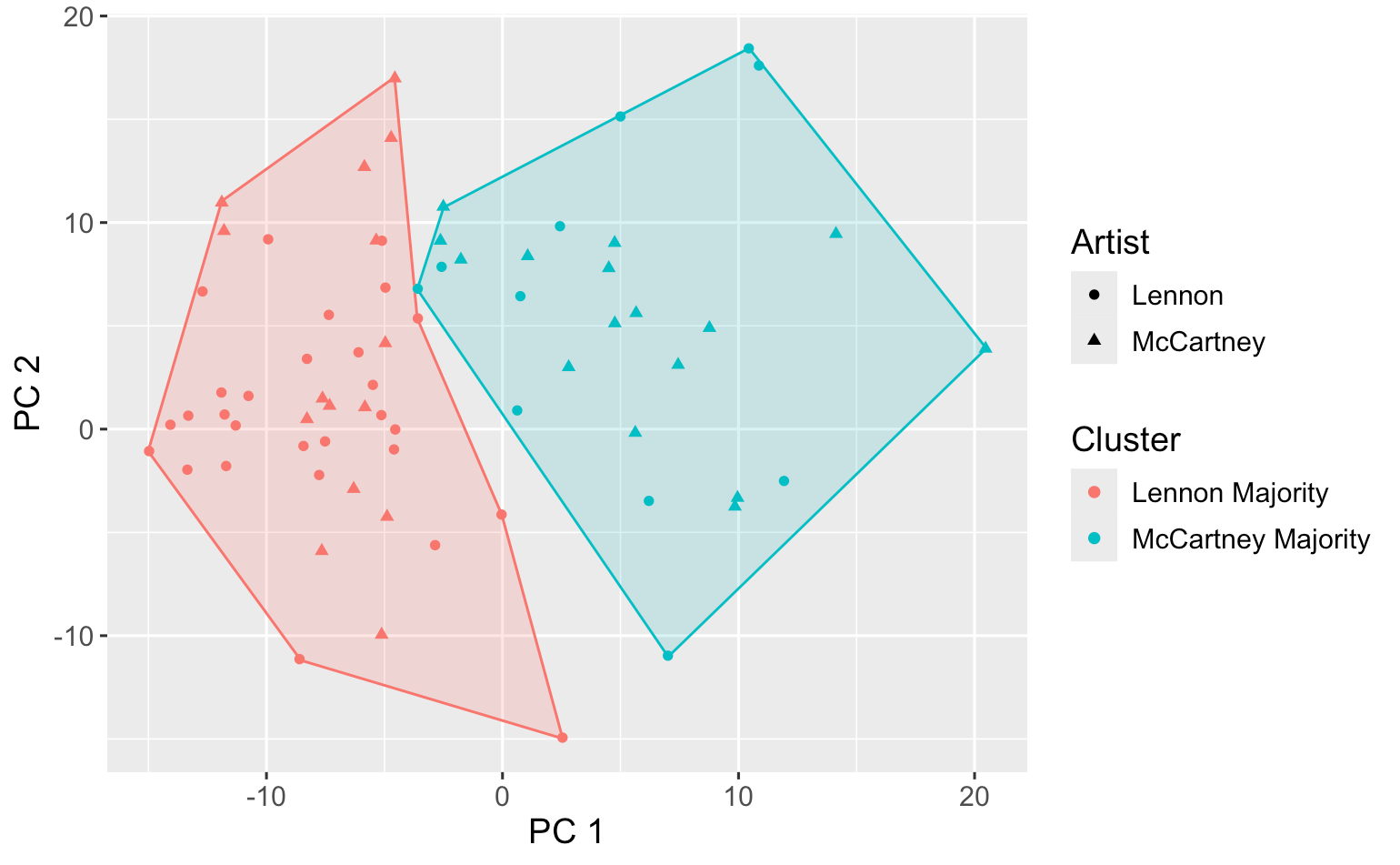}
    \caption{Two-means clustering of Lennon and McCartney songs using the 35 principal components, projected onto the first two dimensions. The algorithm correctly identifies 45 out of 70 songs, noting that the colors are manually assigned using the majority label in each cluster.}
    \label{fig:twomeans}
\end{figure}

As we can see in Figure \ref{fig:twomeans}, we are able to obtain close to 70\% accuracy, which is obtained by assigning each song the majority label of its respective cluster. Upon investigation, the clustering algorithm separates the two groups around $\text{PC}_1 = 0$, and we provide possible insight from this in Section~\ref{disc}.

We also used the embeddings to fit several models that allowed comparing song prediction accuracy. First, we ran a logistic regression with the 35 principal components as the predictors, and performed leave-one-out validation to predict the authorship of all Lennon and McCartney songs. This model is of the form 
\begin{align}
    \text{logit}\l(\mathbb P\l(y_i = 1\r)\r) = \beta_0 + \beta_1\textsf{PC}_1 + \hdots + \beta_{35}\textsf{PC}_{35}.
\end{align}
The leave-one-out prediction accuracy was approximately 72\%, which is close to the accuracy of 75.7\% in \cite{glickman2019data}. Next, we performed a standard KNN classification algorithm, choosing the parameter $k = 5$. This value was chosen for several reasons: first, it aligns with the common practice of choosing $k = \sqrt{n}/2 \approx 4.2$ \citep{boehmke2019hands}, rounding up to avoid ties. Furthermore, it had the optimal out-of-sample accuracy through cross-validation, and it also aligned with the heuristic that each author wrote on average five songs per album. The leave-one-out prediction accuracy for KNN was around 69\%. We finally constructed a random forest model \citep{breiman2001rf},
with $1000$ trees and $6$ predictors sampled for splitting at each node. These two parameters were fine-tuned through the $\texttt{tuneRF}$ function in R, and the resulting leave-one-out prediction accuracy for the random forest was around 66\%. These three models were trained in R through the packages $\texttt{stats}$, $\texttt{class}$, and $\texttt{randomForest}$, respectively.  

\begin{figure}
    \centering
    \includegraphics[width=0.85\linewidth]{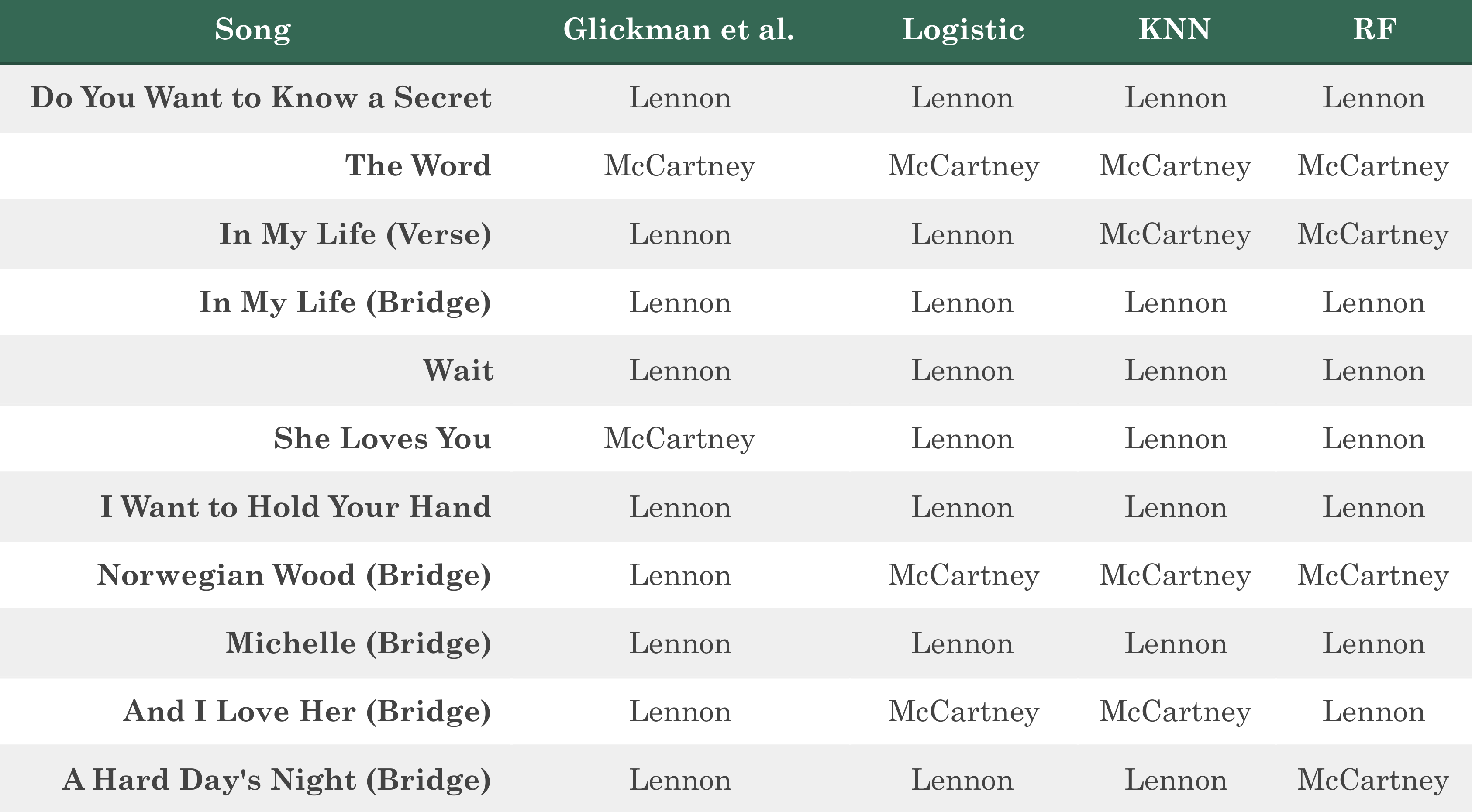}
    \caption{Predicted authorship for a selection of disputed Beatles songs. In over half the cases, all four models produce the same prediction, indicating strong agreement.}
    \label{fig:disputed}
\end{figure}

In Figure \ref{fig:disputed}, we display the author with the highest predicted probability for several songs of disputed authorship. We observe that the \cite{glickman2019data} predictions agree with the logistic regression, KNN, and random forest models on the majority of songs, indicating that the embeddings might be distinguishing genuine musical features between the artists.

\section{Discussion}\label{disc}

The analyses we conducted yield several different findings that offer a fresh perspective on longstanding narratives surrounding the songwriting of The Beatles. Most notably, our figures do not suggest a divergence of style between Lennon and McCartney, but rather a striking \textit{convergence} over time. This challenges the viewpoint of many historical and musicological accounts \citep{pannell2023quant}, which argue that by the mid‑1960s, the Beatles corpus had largely given way to pieces written primarily by only Lennon or only McCartney, with limited input from the partner. Our analysis suggests that their styles actually became increasingly intertwined. 

Among the possible interpretations of this, we may first note that Lennon and McCartney began the Beatles period with distinct musical backgrounds, which naturally led them to write stylistically different songs in the band’s early years. McCartney’s father, James McCartney, was a self‑taught trumpet player and pianist who led an amateur jazz ensemble. He kept a piano in the family home, exposing his son to jazz, music‑hall, and light‑orchestral music from an early age \citep{miles1997paul}. He encouraged McCartney’s musical development by gifting him a trumpet, which McCartney later traded for a Spanish guitar used to compose some of the band's earliest pieces. Lennon, by contrast, was more influenced by early 1950s American rock and roll, combining the blues‑based styles of Elvis Presley and Chuck Berry with the skiffle of Lonnie Donegan popular in Liverpool at the time \citep{harris2013review}. When the two first met in 1957, Lennon was still tuning his five‑string guitar ``like a banjo,'' a technique he had learned from his mother Julia. McCartney later recalled showing him standard guitar chords, and Lennon himself credited McCartney with teaching him to ``play properly'' \citep{beatles2000auto}.

Within this context, it seems natural that their ongoing exchange of ideas would inevitably lead them to shape each other's body of work. As Lennon and McCartney worked side by side, sharing ideas and techniques with one another for hours a day in the studio, their individual approaches grew more aligned as they drew on the strengths and innovations of each other. Our results, showing decreasing distance between album centroids in Figure \ref{fig:album-diffs} and nearly identical growth of within-album variation in Figure \ref{fig:dist-to-cent}, support this idea of mutual influence and experimentation over time. If we also consider that the albums are not considered when creating the principal components, it is surprising to us that their variation over time is so strikingly similar across each album.

\begin{figure}[!t]
    \centering
    \includegraphics[width=0.85\linewidth]{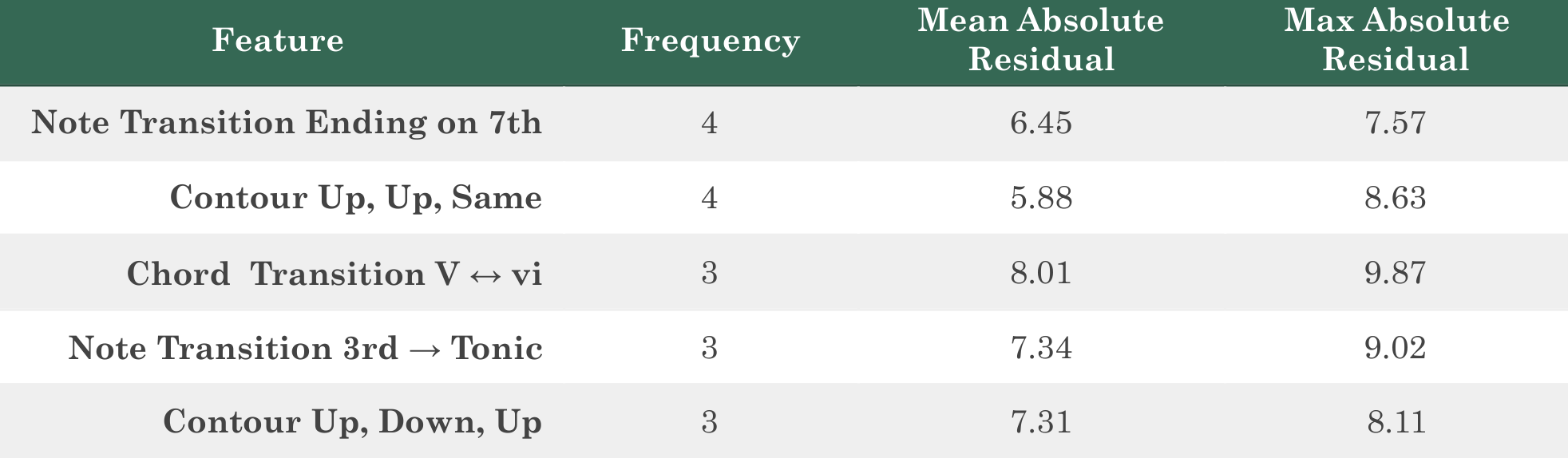}
    \caption{The five features that contributed most to songs being outliers. The frequency refers to how often the feature was among the top ten largest absolute residuals for the six outliers.}
    \label{fig:outlier_features}
\end{figure}

There are also quite a few outliers in the embedding space. After applying the OGK algorithm for outlier detection in high-dimensional data, the six outliers found were: \textit{Please Please Me} and \textit{Tell Me Why} by Lennon, and \textit{Hold Me Tight}, \textit{Things We Said Today}, \textit{Drive My Car}, and \textit{For No One} by McCartney. One way to determine which features contributed the most to these songs being outliers is to calculate the deviance residual contributions 
\begin{align}
    -2\l(x_{ij}\log(p_{ij}) + (1-x_{ij})\log(1 - p_{ij})\r),
\end{align}
where $x_{ij}$ is the $j^{\text{th}}$ binary feature of outlier song $x_i$ ($1 \leq j \leq 137$), and $p_{ij}$ is the predicted probability that $x_{ij} = 1$ as given by our logistic PCA model. 
Figure \ref{fig:outlier_features} illustrates the features that appear most often for the outliers, where the frequency denotes the number of songs for which that feature had a top-ten absolute residual. 

We can see that a note transition ending on the seventh and the contour \textit{up, up, same} appeared in 4 of the 6 outliers as a top feature, meaning these might be what most contributes to these songs being reported as outliers. \textit{Please Please Me} and \textit{Tell Me Why} also have quite large melodic and vocal leaps, with the former being described as ``[Lennon's] attempt at writing a Roy Orbison song'' \citep{lennon2020all}. The other four outliers contain interesting modulations uncommon in The Beatles' early period, some of which are present as features in our dataset and others which are not, such as moving from the major IV to minor iv in \textit{Hold Me Tight}.

Figure \ref{fig:quietbeatle}, which shows the distance from each Harrison song to the Lennon and McCartney centroids from their respective albums, reveals that his songs represented via embeddings are closer to Lennon, and that this distance fluctuates much more with Lennon than it does with McCartney. Overall, Lennon did write songs that were more musically simple than McCartney, and Harrison being a novice songwriter who only really developed this skill in the later years of the Beatles could explain such a connection.

Beyond these descriptive analyses, we demonstrated how logistic PCA facilitates authorship attribution using both unsupervised learning ($k$-means clustering in Figure \ref{fig:twomeans}) and supervised learning (logistic regression, $k$-nearest neighbors, random forest), producing predictions that are generally consistent with more elaborate approaches, as in \cite{glickman2019data}. Such agreement among the models might indicate that there is a true distinction to be made between the musical styles of Lennon and McCartney. 

Our work has shown how performing logistic PCA to create real-valued embeddings from binary features can aid in the statistical analysis of popular music. Applying this method to the early and middle catalog of The Beatles, we have illustrated that embeddings derived from our set of five musical categories can reveal high-level stylistic trends and influence, both over time and across the various songwriters. 

By framing songs as low‑dimensional embeddings, our approach turns a complex musical object into something that can be analyzed with standard multivariate tools, which in turn opens the door to more sophisticated modeling. In particular, once each song is represented as a vector in real space, one can apply time series or dynamic linear models to sequences of songs or albums, allowing prediction of song characteristics and formal inference on how styles differ across writers and time periods. The same embedding framework makes it straightforward to incorporate additional modalities such as rhythm, lyrical sentiment, or audio‑derived features into a joint representation, and to consider songs by many different artists within a single coordinate system. In this sense, embeddings can be viewed as a general strategy to enable a broad class of inferential and predictive analyses in the study of popular music.

\bibliographystyle{plainnat}  
\bibliography{refs} 

\end{document}